\documentclass[10pt,prd,aps,longbibliography,nofootinbib,floatfix, superscriptaddress,preprintnumbers]{revtex4-2}
\usepackage{graphicx}
\usepackage{amsmath}
\usepackage{hyperref}
\usepackage{orcidlink}
\graphicspath{{figures}}
\usepackage{amssymb}

\DeclareMathOperator{\Tr}{Tr}
\newcommand{\orcidauthorDENGLER}{0000-0002-2305-8868}
\newcommand{\orcidauthorMAAS}{0000-0002-4621-2151}
\newcommand{\orcidauthorZIERLER}{0000-0002-8670-4054}

\begin{document}

\title{Scattering of dark pions in Sp(4) gauge theory}

\author{Yannick Dengler \orcidlink{\orcidauthorDENGLER}}
\email{yannick.dengler@uni-graz.at}
\affiliation{University of Graz, Universitätsplatz 5, 8010 Graz, Austria}

\author{Axel Maas \orcidlink{\orcidauthorMAAS}}
\email{axel.maas@uni-graz.at}
\affiliation{University of Graz, Universitätsplatz 5, 8010 Graz, Austria}

\author{Fabian Zierler \orcidlink{\orcidauthorZIERLER}}
\email{fabian.zierler@swansea.ac.uk}
\affiliation{University of Graz, Universitätsplatz 5, 8010 Graz, Austria}
\affiliation{Swansea University, Singleton Park, SA2 8PP Swansea, Wales, United Kingdom}

\begin{abstract}
Analyses of astrophysical data provide first hints on the self-interactions of dark matter at low energies. Lattice calculations of dark matter theories can be used to investigate them, especially in the case of strongly-interacting dark matter. We consider Sp(4) gauge theory with two fundamental fermions as a candidate theory. We compute the scattering phase shift for the scattering of two identical dark pions and determine the parameters of the effective range expansion. Our exploratory results in the supposedly most common interaction channel provide a lower limit for the dark matter mass when compared to astrophysical data. We also provide first benchmarks of velocity-weighted cross-sections in the relevant non-relativistic domain.    
\end{abstract}

\maketitle
\tableofcontents

\section{Introduction}

There exists a large amount of astrophysical data that is not compatible with known physics \cite{Rubin:1978kmz,Bambi:2015mba}. One of the simplest explanations is the existence of a yet unknown type of matter. Despite a substantial amount of effort very little is known to date about the nature of this \textit{dark} matter.

A currently widely debated scenario is self-interacting dark matter (SIDM) \cite{Spergel:1999mh,Kaplinghat:2015aga,Tulin:2017ara,Sagunski:2020spe,Andrade:2020lqq,Eckert:2022qia,Bose:2022obr,Adhikari:2022sbh, Salucci:2018hqu}. One example of such a theory is a hidden confining dark sector. We consider the scenario in which this sector is described by a local Sp(4) gauge group and two degenerate, relatively heavy, flavors of fermions in the fundamental representation. This theory is well-motivated as a SIDM model making use of a $3\to2$ cannibalization process \cite{Hochberg:2014dra, Hochberg:2014kqa} which is known as the Strongly Interacting Massive Particle (SIMP) process\footnote{In an effective description such a $3\to2$ process is inherited by the Wess-Zumino-Witten term \cite{Wess:1971yu, Witten:1983tw}.}.

In addition, at low fermion masses, this theory is a candidate for composite Higgs and other beyond the standard model scenarios \cite{Ferretti:2013kya, Barnard:2013zea,Bennett:2017kga,Cacciapaglia:2020kgq,Bennett:2022yfa, Beylin:2019gtw}. Since both scenarios are relevant for phenomenological studies for physics beyond the Standard Model, the spectral properties of this theory were investigated for a wide range of parameters both in the framework of chiral perturbation theory ($\chi$PT), and on the lattice \cite{Bennett:2019jzz,Bennett:2020qtj,Kulkarni:2022bvh,Bennett:2023rsl,Bennett:2023qwx,Erdmenger:2024dxf}.

However, in view of the lack of any signal of dark matter, these known spectral properties are not sufficient to verify the theory. Dynamical information such as scattering properties is needed\footnote{We note in passing that the same data can be used to constrain form factors relevant to direct detection experiments \cite{Laha:2013gva, Hietanen:2013fya}. These are already experimentally explored in the direct detection experiment CRESST \cite{CRESST:2024ahy}. However, the dependence on the messenger sector, which we do not model, precludes a direct comparison to the present study. Similarly, the electric polarizability can be studied \cite{Drach:2015epq}.}. Scattering properties have been studied in SU(2) gauge theories before \cite{Arthur:2014zda,Drach:2020wux, Drach:2021uhl}. We provide here a first study in this context of Sp(4) gauge theories. The strength of dark matter self-interaction can be constrained from studies of galaxies and galaxy clusters using results of simulated dark matter halos -- see e.g. \cite{Eckert:2022qia, Sagunski:2020spe, Andrade:2020lqq, Kaplinghat:2015aga}. While this astrophysical data still exhibit large systematic uncertainties, these are expected to decrease considerably \cite{Adhikari:2022sbh}. 

Thus, there is a need for such first-principle calculations of dark matter properties. The aim of this work is to provide first lattice-based, non-perturbative predictions that can directly be compared to astrophysical data. We eventually find a lower bound for the dark matter mass of about 100 MeV taking astrophysical data at face value under the assumption that one scattering channel dominates the total scattering cross-section. 

To this end, we start in section~\ref{s:motivation} with a description of this road map, and especially what kind of lattice inputs are needed. For the motivation of the underlying theory which is described in~\ref{s:continuum}, as well as its spectral properties, we refer to the literature \cite{Bennett:2019jzz,Bennett:2023rsl,Hochberg:2014dra, Hochberg:2014kqa,Kulkarni:2022bvh}. In section~\ref{s:lattice} we describe the details of the lattice calculation to obtain the inputs identified in section~\ref{s:motivation}. Readers not interested in the technical details can skip this section. We combine the lattice calculations with the road map discussed in section~\ref{s:motivation} in section~\ref{s:results}, delivering our final results. We summarize the paper with a short discussion and conclude in section~\ref{s:conclusion}. Preliminary results have been presented in \cite{Zierler:2022qfq,Dengler:2023szi}.

\section{Motivation}\label{s:motivation}

In the current absence of a direct detection of dark matter, astrophysical simulations of dark matter provide valuable input to constrain its properties. Recent investigations provided upper bounds on the self-interaction cross-section from various sources as low as $\sigma/m<0.13$ cm$^2$/g (with possible systematic errors of order $0.1$ cm$^2$/g) \cite{Andrade:2020lqq}, $\sigma/m<0.19$ cm$^2$/g \cite{Eckert:2022qia}, and $\sigma/m<0.35$ cm$^2$/g \cite{Sagunski:2020spe}. The typical velocities in these systems are of order $\mathcal O (1000~{\rm km/s})$, the escape velocity of galaxies. The velocity weighted cross-section $\left<\sigma v\right>$ was investigated in \cite{Kaplinghat:2015aga, Sagunski:2020spe}, and it was found that their results prefer a mild velocity dependence.

In the following, we present how to arrive at this quantity from first-principle calculations from lattice field theory. We start by defining the velocity-weighted cross-section
\begin{align}
\begin{split}
\left<\sigma v\right> &= \int_0^{v_{esc}} v\, \sigma(v)\, f(v, \langle v\rangle)\, dv, \\
f(v) &= \frac{32 v^2}{\pi^2 \left<v\right>^3}\exp\left(-\frac{4 v^2}{\pi \left<v\right>^2}\right). \\
\label{eq: sigma v}
\end{split}    
\end{align}

\noindent Here, $v$ is the relative velocity between two dark matter particles, which we describe by a Maxwell-Boltzmann distribution with mean velocity $\langle v\rangle$\footnote{It can be shown that the distribution of the relative velocities between two particles, each described by a Maxwell-Boltzmann distribution, again follows a Maxwell-Boltzmann distribution where the mass is substituted by the reduced mass of the two particles. For equal masses this means that the mean relative velocity is $\sqrt{2}$ times the mean velocity of a single particle.}\cite{Tulin:2013teo}, $v_{esc}$ is the escape velocity of the halo which we assume to be much larger than $\langle v\rangle$ and $\sigma$ is the cross-section. One finds that $\left<\sigma v\right>$ depends solely on the cross-section as a function of $v$. It is useful to decompose the cross-section in terms of partial wave amplitudes,
\begin{align}\label{eq:sigma_def}
    \sigma&=\int d\Omega \frac{d\sigma}{d\Omega}, \nonumber \\ 
    \frac{d\sigma}{d\Omega}&=\frac{1}{\left(16\pi\right)^2}\frac{1}{p^2}|{\cal M}|^2,\\
    {\cal M}&=16\pi\sum_l(2l+1)e^{i\delta_l}\sin(\delta_l)P_l(\cos\theta),\nonumber
\end{align}
where the last lines show the decomposition of the matrix element $\cal{M}$ in partial waves. Here, $d\Omega = \sin(\theta) d\theta d\phi$ is the solid angle, $p$ is the relative momentum in the center-of-mass frame and $\delta_l$ is the scattering phase-shift of angular momentum quantum number $l$. It is useful to introduce the cross-section of the different partial waves as
\begin{equation}\label{eq:sigma(l)}
    \sigma_l(p) = 4\pi(2l+1)\frac{1}{p^2}\sin^2(\delta_l).
\end{equation}
At small momenta above the elastic threshold the phase shift $\delta_l$ can be described by an effective range expansion (ERE)\footnote{There are different sign conventions for the scattering length in the literature. In this paper, we are working with the convention used in \cite{Sakurai:2011zz, Kondo:2022lgg}. In contrast, other papers like \cite{Jenny:2022atm, Arthur:2014zda, Bijnens:2011fm} use the opposite sign convention for the leading coefficient $a_l$.}
\begin{equation}
    \label{eq:ERT}
    p^{2l+1}\cot{\delta_l} = -\frac{1}{a_l^{2l+1}}+\frac{p^2}{2r_l^{2l-1}}+\mathcal{O}(p^4)
\end{equation}
This expansion is valid for $r_l p \lesssim 1$, which is always fulfilled for typical velocities of dark matter in galaxies. We consider here ERE for s-wave scattering to second order which includes the scattering length $a_0$ and the \textit{effective range} $r_0$.  The s-wave ($l=0$) cross-section using ERE then reads
\begin{equation} \label{eq:ERT_sigma}
\sigma_0(p) = \frac{4\pi a_0^2}{|1-\frac{a_0 r_0}{2}p^2+ipa_0|^2}.
\end{equation}
We can relate the momentum and the mass to a velocity via $p = m v \gamma$, where $\gamma$ is the Lorentz factor. This framework allows us to fit the discrete data points of phase shifts and momenta that we get from the lattice to obtain the cross-section as a function of the velocity. As can be seen from \eqref{eq: sigma v}, this is sufficient to calculate the velocity-weighted cross-section.

\noindent Note that in Ref.~\cite{Kondo:2022lgg} the aforementioned results on the velocity-dependent cross-section from astrophysical data from \cite{Kaplinghat:2015aga} have been used to determine the best-fit coefficients of the ERE. In contrast, we will determine these parameters from a first principle lattice simulations for a candidate theory. They can be compared to this fit, to understand (possible) deviations of our candidate theory from the optimal, yet unknown, true theory of dark matter.

\section{The continuum theory}\label{s:continuum}

The Lagrangian of Sp(4) gauge theory in isolation with two mass-degenerate fundamental Dirac fermions with a bare mass $m_0$ in Minkowski spacetime is given by
\begin{align}
    \mathcal L = -\frac{1}{2} \Tr F_{\mu\nu} F^{\mu\nu} + \bar u  \left( i \gamma^\mu D_\mu - m_0 \right) u + \bar d  \left( i \gamma^\mu D_\mu - m_0 \right) d.
\end{align}
In analogy to QCD we denote the two fermion fields as $u$ and $d$. Here $F_{\mu\nu}$ is the field strength tensor of Sp(4) gauge theory and $D$ is the covariant derivative involving the gauge fields in the fundamental representation as 
\begin{align}
    D_\mu u = \partial_\mu u + i g A_\mu u = \partial_\mu u + i g A_\mu^a \tau^a u 
\end{align}
Due to the pseudoreality of the generators\footnote{This means that there is a color matrix $S$ for which $S \tau^a S = (\tau^a)^T$ for all generators $\tau^a$.}, the global symmetries of the Lagrangian are enlarged in comparison to QCD-like theories with a complex representation \cite{Kosower:1984aw}. The massless Lagrangian is invariant under a global U(1)$_A \times$ SU(4) acting on the fermion fields. The $U(1)_A$ is broken by the axial anomaly and the global SU(4) is broken spontaneously by the fermion condensate and explicitly by non-vanishing fermion masses down to a global Sp(4) symmetry.

Spontaneous chiral symmetry breaking gives rise to the SU(4)/Sp(4) coset and thus five light pseudo-Nambu-Goldstone bosons (pNGBs) \cite{Kogut:2000ek}, which are the DM candidates referred to as \textit{dark pions}. Our goal is to determine their scattering cross-sections as outlined in section~\ref{s:motivation}. 

The scattering of two pions can occur in three different channels classified under remaining global Sp(4) symmetry. Mesons that consist out of two fundamental fermions are either in a 5-dimensional representation, a 10-dimensional representation, or singlet representation. This follows from the decomposition of the product of two four-dimensional fundamental representations into its irreducible representations (irreps) as \cite{Drach:2017btk, Feger:2019tvk}
\begin{align}
    4 \otimes 4 = 1 \oplus 5 \oplus  10. 
\end{align}
The pions live in the $5$-dimensional irrep of Sp(4) whereas the vector mesons live in the 10-dimensional irrep. For two-pion scattering the irreps of the two-pion product state are relevant which are given by
\begin{align}
    5 \otimes 5 = 1 \oplus 10 \oplus 14.
\end{align}
Pions will scatter in each of these three channels, which need to be calculated separately on the lattice. We can identify these channels with their equivalent processes in QCD. The scattering channel involving the $10$-dimensional representation and the one-dimensional representation can contain other single-mesons states. Borrowing the QCD nomenclature, these resonances occur e.g. in processes such as $\pi\pi \to \rho$ for the $10$-dimensional irrep and $\pi\pi \to \sigma/f_0$ for the singlet representation. In QCD, those would be the isospin $I=1$ and isospin $I=0$ channels. In this paper, we focus on the $14$-dimensional representation which can be shown to involve (among other equivalent processes) the scattering of two identical pions \cite{Arthur:2014zda} such as $\pi^+ \pi^+$, and thus corresponds to the isospin $I=2$ channel in QCD. It arguably contributes most to the total cross-section since 14 out of the total 25 combinations of pions scatter in this channel.

\section{Lattice setup}\label{s:lattice}

We discretize the Euclidean action using the standard Wilson plaquette action and two mass-degenerate unimproved Wilson Dirac Fermions \cite{Wilson:1974sk} on hypercubic lattices with a volume of  $L^3 \times T = a^4~N_L^3 \times N_T$. The gauge configurations were generated using the Hybrid Monte Carlo (HMC) \cite{Duane:1987de} algorithm. We perform measurements on the configurations with the HiRep code \cite{DelDebbio:2008zf, HiRepSUN} which has been extended to symplectic gauge groups \cite{HiRepSpN}. An overview of the ensembles characterized by the inverse gauge coupling $\beta = 8/g^2$ and the bare fermion masses $m_0$ is given in table~\ref{t:ensembles}. The ensembles have been partially generated for earlier spectroscopic investigations in \cite{Bennett:2019jzz, Kulkarni:2022bvh, Bennett:2023wjw}. We supplemented these ensembles by new ones where needed such that for every set of bare parameters $(\beta, am_0)$, we have at least three different lattices sizes $N_L$ and thus at least three different lattice momenta.

\begin{table}
    \caption{All ensembles used in this work. They are defined by the inverse gauge coupling $\beta$, the bare fermion masses of the degenerate fermions $a m_{0}$ in lattice units, the lattice extent ($N_L \times N_T$) and the number of configurations $n_{\rm config}$. We further give the ratio of the pseudoscalar meson mass to the vector meson mass $m_\pi/m_\rho$ as well as the ground state energy in lattice units for the one-pion channel $m_\pi$ and the two-pion channel $E_{\pi\pi}.$ For comparison with the leading-order $\chi$PT prediction we report the perturbatively renormalized pion decay constant $f_\pi$.}
    \label{t:ensembles}
    \begin{tabular}{|l|l|r|r|r|l|l|l|l|l|}
        \hline
        $\beta$ & $a m_{0}$ & $N_L$ & $N_T$ & $n_{\rm config}$ & $ m_\pi/m_\rho$ & $a m_\pi$ & $a E_{\pi\pi}$ & $a f_{\pi}$ & $\langle P \rangle$ \\ \hline \hline
        6.9 & -0.87 & 10 & 20 & 976 & 0.8744(43) & 0.7425(12) & 1.4961(22)  & 0.1313(25) & 0.550680(46) \\
        6.9 & -0.87 & 12 & 24 & 400 & 0.8754(41) & 0.7414(15) & 1.4891(27)  & 0.13195(91) & 0.550441(54) \\
        6.9 & -0.87 & 16 & 32 & 100 & 0.8762(28) & 0.74060(96) & 1.4820(23)  & 0.1324(35) & 0.550525(61) \\ \hline
        6.9 & -0.9 & 8 & 16 & 651 & 0.795(11) & 0.6241(25) & 1.2799(48)  & 0.1014(44) & 0.557959(99) \\
        6.9 & -0.9 & 8 & 24 & 402 & 0.811(12) & 0.6267(26) & 1.2806(53)  & 0.0969(20) & 0.55796(10) \\
        6.9 & -0.9 & 10 & 20 & 1273 & 0.7998(38) & 0.5738(12) & 1.1602(26)  & 0.1044(21) & 0.557172(40) \\
        6.9 & -0.9 & 12 & 24 & 2904 & 0.8110(22) & 0.56409(54) & 1.1339(14)  & 0.10484(87) & 0.557009(18) \\
        6.9 & -0.9 & 14 & 24 & 942 & 0.8115(27) & 0.56222(63) & 1.1280(16)  & 0.10599(58) & 0.556981(26) \\
        6.9 & -0.9 & 16 & 32 & 546 & 0.8156(28) & 0.56275(57) & 1.1283(12)  & 0.1064(13) & 0.556921(25) \\
        6.9 & -0.9 & 18 & 36 & 356 & 0.8135(24) & 0.56121(58) & 1.1245(12)  & 0.10576(91) & 0.556987(24) \\ \hline
        6.9 & -0.91 & 12 & 24 & 1268 & 0.7698(77) & 0.4920(10) & 0.9950(27)  & 0.0949(13) & 0.559351(28) \\
        6.9 & -0.91 & 14 & 24 & 513 & 0.7756(81) & 0.4857(12) & 0.9781(29)  & 0.0945(23) & 0.559409(34) \\
        6.9 & -0.91 & 16 & 32 & 435 & 0.7658(65) & 0.48610(86) & 0.9765(19)  & 0.0948(23) & 0.559353(27) \\ \hline
        6.9 & -0.92 & 12 & 24 & 63 & 0.738(61) & 0.416(19) & 0.885(14)  & 0.0853(68) & 0.56145(14) \\
        6.9 & -0.92 & 14 & 24 & 550 & 0.699(10) & 0.3926(14) & 0.7914(40)  & 0.0724(19) & 0.562096(34) \\
        6.9 & -0.92 & 16 & 32 & 176 & 0.670(11) & 0.3894(14) & 0.7848(35)  & 0.0821(15) & 0.562116(42) \\
        6.9 & -0.92 & 24 & 32 & 467 & 0.7035(31) & 0.38649(51) & 0.7734(12)  & 0.08260(35) & 0.562077(14) \\ \hline
        7.05 & -0.835 & 8 & 24 & 402 & 0.777(13) & 0.6585(60) & 1.345(15)  & 0.0481(47) & 0.577085(74) \\
        7.05 & -0.835 & 12 & 24 & 313 & 0.790(11) & 0.4616(15) & 0.9424(32)  & 0.0792(14) & 0.575237(39) \\
        7.05 & -0.835 & 14 & 24 & 619 & 0.7877(91) & 0.4417(17) & 0.9085(30)  & 0.0793(20) & 0.575368(25) \\
        7.05 & -0.835 & 20 & 36 & 100 & 0.7945(61) & 0.4380(10) & 0.8792(27)  & 0.0796(31) & 0.575269(29) \\ \hline
        7.05 & -0.85 & 12 & 24 & 84 & 0.611(33) & 0.3778(57) & 0.786(22)  & 0.0582(39) & 0.577835(69) \\
        7.05 & -0.85 & 14 & 24 & 167 & 0.716(26) & 0.3496(25) & 0.7236(67)  & 0.0675(17) & 0.577429(44) \\
        7.05 & -0.85 & 16 & 32 & 101 & 0.660(17) & 0.3375(17) & 0.6892(41)  & 0.0669(11) & 0.577413(41) \\
        7.05 & -0.85 & 24 & 36 & 100 & 0.7118(64) & 0.33076(97) & 0.6638(23)  & 0.0684(19) & 0.577371(24) \\ \hline
        7.2 & -0.78 & 8 & 24 & 401 & 0.8770(81) & 0.8089(42) & 1.617(11)  & 0.0402(50) & 0.590527(59) \\
        7.2 & -0.78 & 10 & 20 & 195 & 0.648(16) & 0.5508(48) & 1.1345(88)  & 0.0497(20) & 0.589788(65) \\
        7.2 & -0.78 & 12 & 24 & 150 & 0.835(19) & 0.4382(34) & 0.9024(84)  & 0.0569(14) & 0.589547(56) \\
        7.2 & -0.78 & 14 & 24 & 425 & 0.7762(83) & 0.3857(14) & 0.7951(35)  & 0.06569(73) & 0.589362(26) \\
        7.2 & -0.78 & 16 & 32 & 265 & 0.7930(90) & 0.3809(11) & 0.7703(31)  & 0.0645(11) & 0.589253(22) \\
        7.2 & -0.78 & 24 & 36 & 508 & 0.7852(30) & 0.36963(39) & 0.74360(79)  & 0.0646(26) & 0.5892779(85) \\ \hline
        7.2 & -0.794 & 12 & 24 & 101 & 0.732(26) & 0.3932(63) & 0.823(13)  & 0.0389(24) & 0.590837(54) \\
        7.2 & -0.794 & 14 & 24 & 234 & 0.691(31) & 0.3234(26) & 0.6888(66)  & 0.0533(14) & 0.590422(39) \\
        7.2 & -0.794 & 16 & 32 & 101 & 0.796(27) & 0.3097(17) & 0.6463(50)  & 0.0570(13) & 0.590330(40) \\
        7.2 & -0.794 & 28 & 36 & 504 & 0.7163(57) & 0.28524(35) & 0.57582(97)  & 0.05689(71) & 0.5904516(67) \\ \hline
    \end{tabular}
\end{table}

\subsection{Interpolating operators}
In order to extract the scattering properties in the isospin $I=2$ channel, we need to compute the energy difference between the interacting and non-interacting pions in a finite volume -- see also section~\ref{s:Lüscher}. Thus, we need to compute the energies of both one-pion and two pion states. For this, we consider the local one-pion and two-pion interpolators
\begin{align} 
    \mathcal O_{\pi}(x) &=  \bar u(x) \gamma_5 d(x), \label{eq:two-point-pi} \\
    O_{\pi\pi}(x) &= O_{\pi}(x)O_{\pi}(x) = \bar u(x) \gamma_5 d(x) \bar u(x) \gamma_5 d(x), \label{eq:two-point-pipi} \\
   \mathcal O_{\gamma_0 \gamma_5}(x) &=  \bar u(x) \gamma_0 \gamma_5 d(x), \label{eq:two-point-pi-decay}
\end{align}
and calculating the two-point correlation function projected to zero momentum 
\begin{align}\label{eq:correlator}
    C_\mathcal{O}(t-t') = \sum_{\vec x, \vec y} \langle \bar{\mathcal O}(\vec x, t) \mathcal{O}(\vec y, t') \rangle.
\end{align}
By formally inserting the complete set of states of the Hamiltonian, the correlator can be written as
\begin{align}\label{eq:correlator_expansion}
    C_\mathcal{O}(t) = \sum_n \frac{1}{2E_n} \langle 0 \vert \bar{\mathcal O} \vert n \rangle \langle n \vert \mathcal O \vert 0 \rangle e^{-E_n t}.
\end{align}
At large Euclidean times the correlator will be dominated by the ground state energy, assuming that a suitable operator basis with a sufficient overlap was chosen. If the excited state contributions decay quickly enough, the ground state energy can be extracted by fitting the correlator's exponential decay at large $t$. We check that this dominance of the ground state is indeed the case, by visually inspecting an effective mass $m_{\rm eff}(t)$ implicitly defined on a lattice with temporal extent $T$ as 
\begin{align}
    \frac{C(t-1)}{C(t)} =  \frac{e^{-m_{\rm eff}(t)\cdot(t-1)} \pm e^{-m_{\rm eff}(t)\cdot(T-(t-1))}}{ e^{-m_{\rm eff}(t)\cdot t} \pm e^{-m_{\rm eff}(t)\cdot(T-t)}} ,
\end{align}
which will show a plateau if the behavior of the correlator is well approximated by a single exponential term. The sign is chosen to be positive for symmetric correlators and negative for antisymmetric correlators. 

In order to compare our results with the predictions of leading order $\chi$PT, we determine the pion decay constant from the $\mu=0$ axial-vector operator \eqref{eq:two-point-pi-decay}. The unrenormalized decay constant $f_\pi^0$ is defined through the corresponding matrix-element\footnote{We chose a convention that corresponds to $f_\pi \approx 93 {\rm MeV}$ for the QCD pion.}. Thus, at large Euclidean times the correlator Eq.~\eqref{eq:correlator} on a lattice with finite temporal extent $T$ of Eq.~\eqref{eq:two-point-pi-decay} has the form
\begin{align}\label{eq:pion_decay_constant}
    \lim_{t\to\infty} C_{\mathcal{O}_{\gamma_0 \gamma_5}}(t) = \frac{1}{2m_\pi} \vert \langle 0 \vert \mathcal O_{\gamma_0 \gamma_5} \vert {\rm PS} \rangle \vert^2 \left(  e^{-m_\pi t} + e^{-m_\pi (T-t)}  \right) = \frac{\left( f^0_\pi \right)^2 m_\pi}{2} \left(  e^{-m_\pi t} + e^{-m_\pi (T-t)}  \right).
\end{align}
For the purposes of this investigation we follow the approach of \cite{Bennett:2017kga,Kulkarni:2022bvh} and estimate the renormalization constant $Z_A$ needed for the renormalized pion decay constant $f_\pi = Z_A f_\pi^0$ using leading order lattice perturbation theory \cite{Martinelli:1982mw}. We stress, that this is only used for comparing to the effective field theory predictions and has no effect on our determination of the scattering properties.

The Wick-contractions for the correlators defined by \eqref{eq:correlator} with the lattice sites $n$ and $m$ made explicit,  can be represented diagrammatically as \cite{Arthur:2014zda}
\begin{align}
    \langle  \bar{\mathcal O}_{\pi}(n) \mathcal O_{\pi}(m)  \rangle_F &= \vcenter{\hbox{\includegraphics[scale=0.5,page=4]{contractions.pdf}}}, \\
    \langle \bar{\mathcal  O}_{\pi\pi}(n) O_{\pi\pi}(m)  \rangle_F &= \vcenter{\hbox{\includegraphics[scale=0.5,page=27]{contractions.pdf}}} - \vcenter{\hbox{\includegraphics[scale=0.5,page=26]{contractions.pdf}}},
\end{align}
where solid lines denote the fermion propagator with suppressed color and spin indices $D(n|m)^{-1}$ between the lattice sites $n$ and $m$. 

We use $Z_2\times Z_2$ stochastic noise sources with spin-dilution \cite{Foley:2005ac} for the inversion of the Dirac operator. For the pion correlator we always choose four different sources to calculate the one-to-all propagator. For the two-pion operators we use $4 \leq n_{\rm src} \leq 16 $ sources. We find that the number of sources used has no quantitative impact within statistical uncertainties on the extracted energies.

The two-pion correlator can receive a constant contribution to the sum of exponentials in \eqref{eq:correlator_expansion}. This is a generic feature for scattering states of multiple hadrons \cite{Kim:2003xt, Prelovsek:2008rf} due to the two pions propagating in opposite directions on the periodic lattice. A similar behavior has also been observed for simulations at finite temperature in the high temperature phase -- see Ref.~\cite{Umeda:2007hy} for a detailed discussion and a comparison of different techniques to remove this contribution. This constant is indeed found to be sizable in our correlation functions. We therefore remove this constant contribution by performing a numerical derivative on the correlator $C(t)$,
\begin{align}
    \tilde C(t) = \frac{C(t-1)-C(t+1)}{2}.
\end{align}
Note, that this changes the correlator from a symmetric quantity with respect to the lattice midpoint $T/2$ to an antisymmetric one. 

For the fitting of the correlator, we use the corrfitter package to extract energy levels \cite{peter_lepage_2021_5733391}. It allows us to keep the fit stable using a Bayesian approach. We typically exclude the first three data point in Euclidean time $t$ and fit the remaining correlator to several exponential terms. For the results shown here we only consider one operator for the two-pion channel. To improve the systematics and to get more data points in the Lüscher analysis, one could include more operators with non-vanishing momentum and/or smeared operators in the respective channel and perform a variational analysis by solving the generalized eigenvalue problem. In light of the exploratory nature of this paper, we defer this to future work.

For the Lüscher analysis, we need the energy of two non-interacting pions. This energy is given by twice the pion mass in the infinite volume limit. We determine it by fitting the masses obtained in a finite spatial volume $L^3$ to the function
\begin{align}\label{eq:infinite_volume}
    m_\pi(L) = m^\infty_\pi \left( 1 + A \frac{e^{-m^\infty_\pi L }}{\left( m^\infty_\pi L \right)^{-3/2}} \right).
\end{align}
This form of the fit function is motivated by chiral lattice perturbation theory. Within this framework the coefficient $A$ is not a free parameter. However, we choose to follow \cite{Bennett:2019jzz} and treat $A$ as a free parameter to account for deviations from $\chi$PT. The results from the fits are presented in table~\ref{t:results}.

\subsection{Lüscher analysis}
\label{s:Lüscher}

\begin{figure}
\begin{center}
    \includegraphics[width=0.9\textwidth]{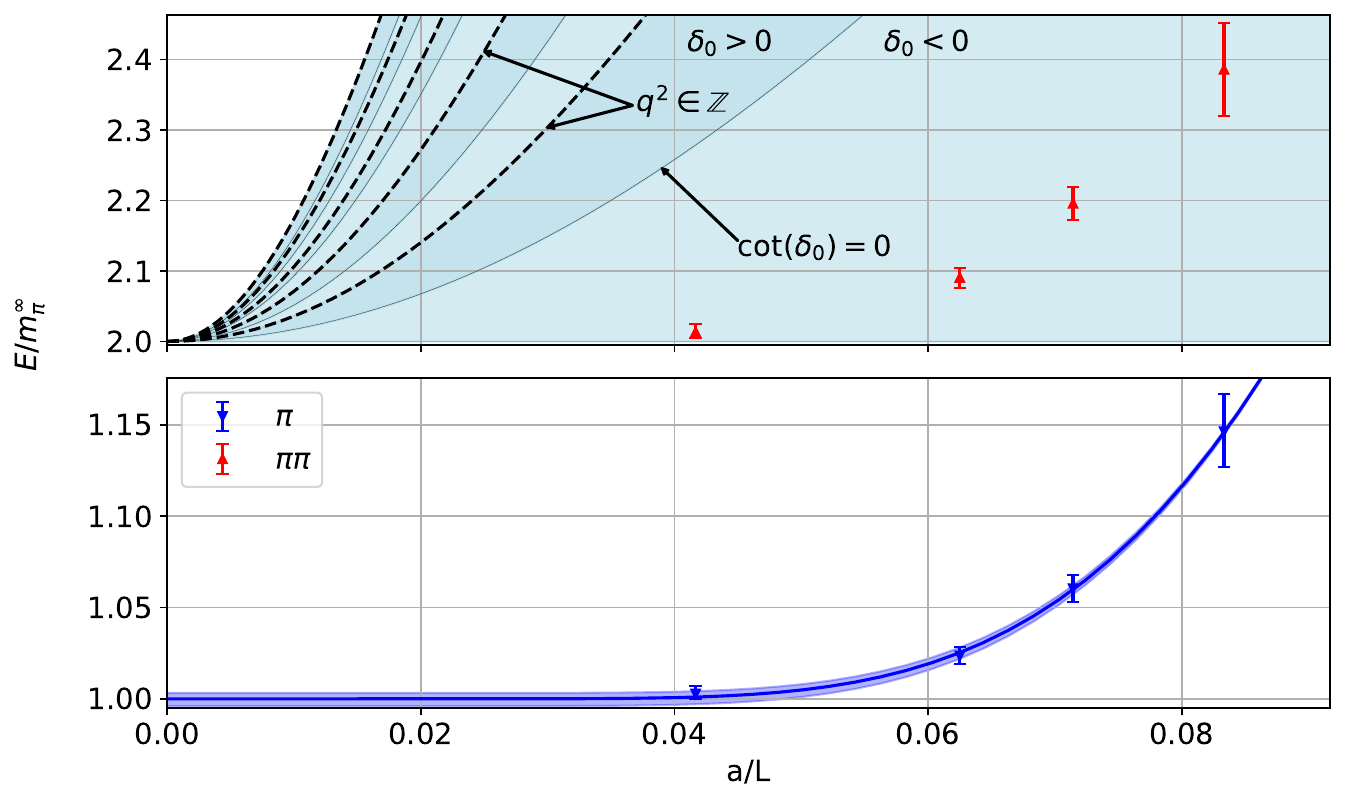}
    \caption{Bottom panel: The extracted pion energy levels plotted against the unitless inverse spacial lattice extent for the ensemble with $\beta = 7.05$ and $am_0 = -0.85$. The line and band indicate the mean and estimated error of the fit using \eqref{eq:infinite_volume}. The results of the fit for all ensembles is given in table~\ref{t:results}. The y-axes in both panels are scaled by the result of the fit. Top panel: Same plot but for the two-pion energies. The dashed lines are the trivial energy levels obtained for integer values of $q^2$. The blue light (dark) shaded region mark where the Lüscher formalism results in negative (positive) values for $p\,\cot(\delta_0)$ \eqref{eq:tan_PS}. The regions are obtained by solving for zeros of the Lüscher Zeta function.}
    \label{plot:E_vs_L_inv}
\end{center}
\end{figure}

Interactions between particles shift the finite volume energy levels compared to the continuum. The Lüscher analysis \cite{Luscher:1985dn, Luscher:1986pf, Luscher:1990ux} is a tool to relate these energy shifts to infinite volume scattering properties. We employ here two particles with vanishing total momentum only. The formalism is valid for energy levels between the elastic threshold and the first inelastic threshold (presumed in our case to be $2m_\pi^\infty < E_{\pi\pi} < 4m_\pi^\infty$). Consequently, we exclude all data points from table~\ref{t:ensembles} outside this elastic region. The momentum can be calculated from the energy levels using a dispersion relation that accounts for the finite lattice spacing

\begin{align}
\begin{split}
    \cosh\left(\frac{aE_{\pi\pi}}{2}\right) = \cosh(am_{\pi}^\infty)+2\sin\left(\frac{ap}{2}\right)^2
    \label{eq:dispersion_relation}
\end{split}
\end{align}
though it turns out that the continuum version ($E_{\pi\pi}=2\sqrt{m^{\infty 2}_{\pi}+p^2}$) works just as well\footnote{This is expected, as the two dispersion relations are identical in the limit of $p\to 0$.}. After ensuring that the results do not depend on the dispersion relation used, we employ the continuum one. We define the generalized momentum $q = p\frac{L}{2\pi}$ which allows the calculation of the phase shift $\delta_0$. In our case for vanishing total momentum the phase shift is connected to the generalized momentum by the following formula that includes the transcendental Zeta function $\mathcal{Z}$
\begin{align}
\begin{split}
   \tan(\delta_0(q))=\frac{\pi^{\frac{3}{2}}q}{\mathcal{Z}^{\Vec{0}}_{00}(1,q^2)}.
   \label{eq:tan_PS}
\end{split}
\end{align}
For the calculation of $\mathcal{Z}$, we refer to \cite{Rummukainen:1995vs,Jenny:2022atm}. One can immediately see, that the sign of $\tan(\delta_0)$ is exclusively determined by the sign of $\mathcal{Z}^{\Vec{0}}_{00}(1,q^2)$. This can be used to determine regions of positive (negative) values of $\tan(\delta_0)$ from the values of $E_{\pi\pi}(L)$ directly as is shown in figure~\ref{plot:E_vs_L_inv} for one of our ensembles. The regions are separated by choosing integer values and zeros in the Zeta function for $q^2$. In the limit of vanishing momentum this sign also coincides with the opposite sign of the scattering length in Eq.~\eqref{eq:ERT}. 

\section{Results}\label{s:results}

\begin{figure}
\begin{center}
    \includegraphics[width = 0.90\textwidth]{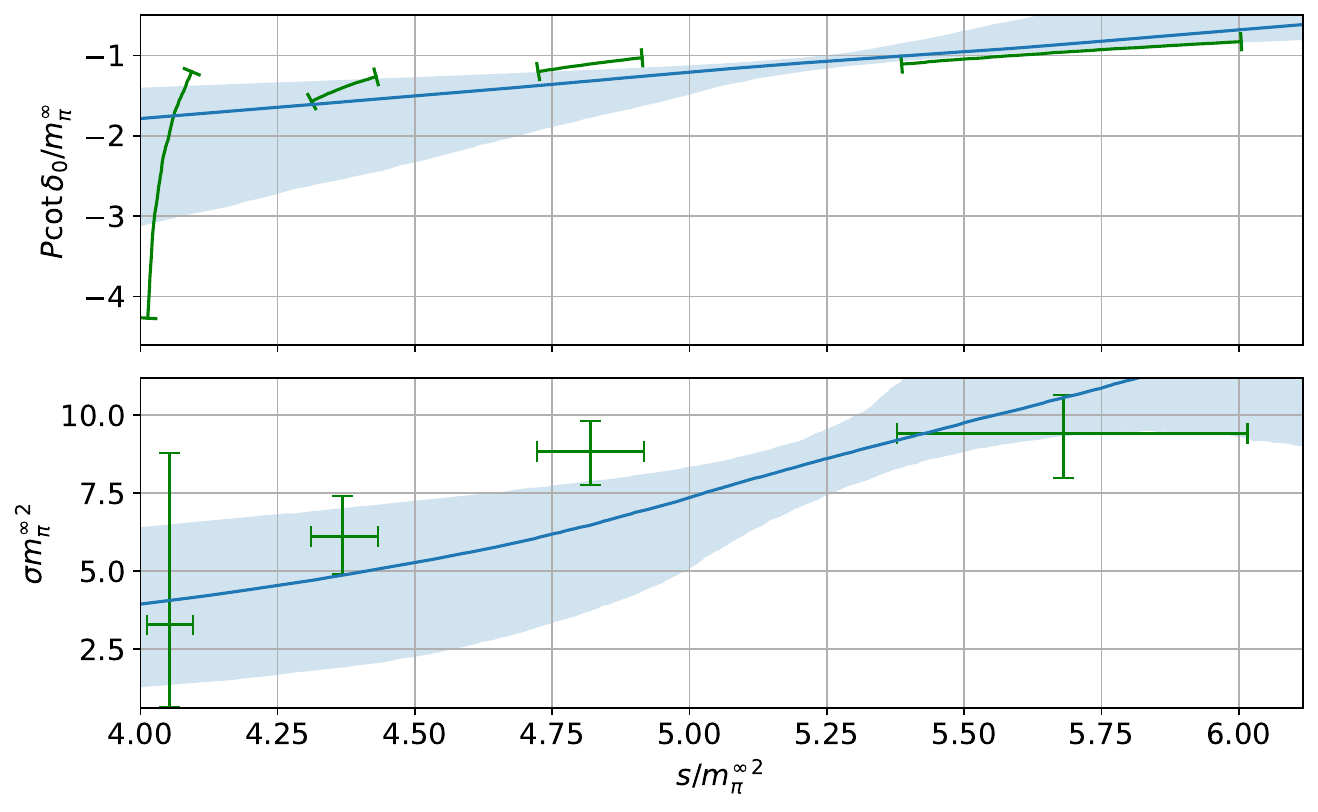}
    \caption{Top panel: Result of \eqref{eq:tan_PS} plotted against the center-of-momentum energy squared for the same ensemble as in fig~\ref{plot:E_vs_L_inv} ($\beta = 7.05$ and $am_0 = -0.85$). The blue line and band show the result of fitting the data to \eqref{eq:ERT}. The curved error bars indicate, that the two axes are not independent and only values on the curved lines are allowed. Bottom panel: Result of using the data points above in \eqref{eq:ERT_sigma} to obtain the s-wave cross-section.}
    \label{plot:p_cot_PS}
\end{center}
\end{figure}
\begin{figure}
\begin{center}
    \includegraphics[width=0.9\textwidth]{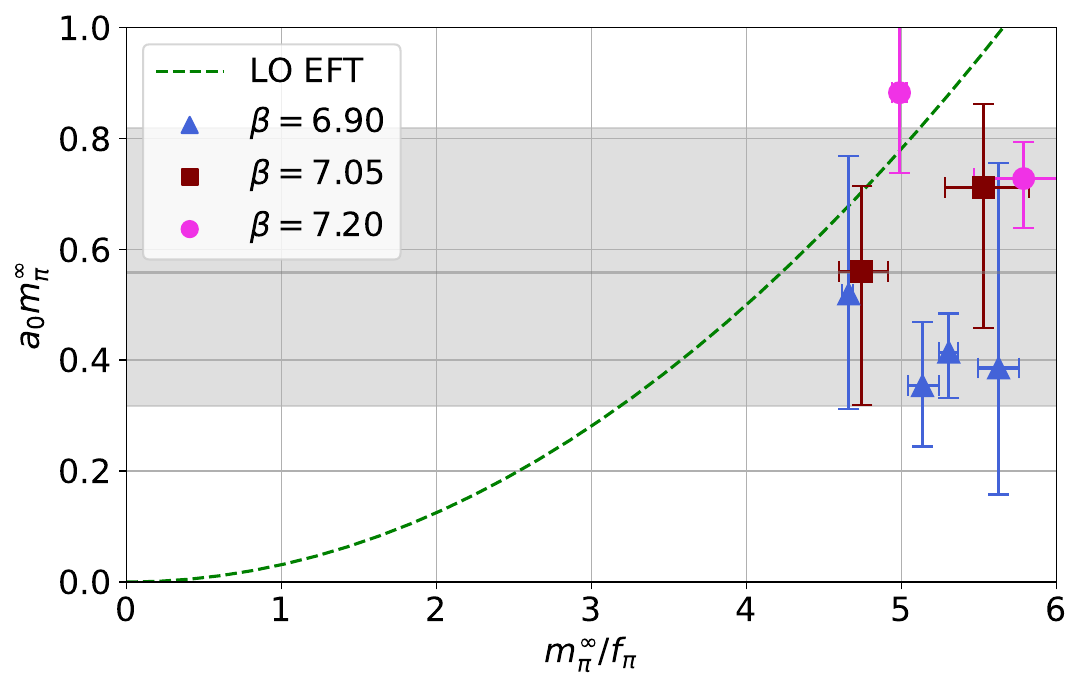}
    \caption{The scattering length obtained in figure~\ref{plot:p_cot_PS} plotted against the ratio of the mass and the decay constant of the pion. Different colors and symbols correspond to different values for the inverse coupling $\beta$. We observe a consistent positive scattering length across all ensembles. The horizontal gray line and band indicate a central value and error for the scattering length estimated using all of our ensembles. The green dashed line shows the expected result from leading order $\chi$PT \cite{Bijnens:2011fm} (Note the different sign convention).}
    \label{plot:scattering_length_fpi}
\end{center}
\end{figure}
\begin{figure}
\begin{center}
    \includegraphics[width=0.9\textwidth]{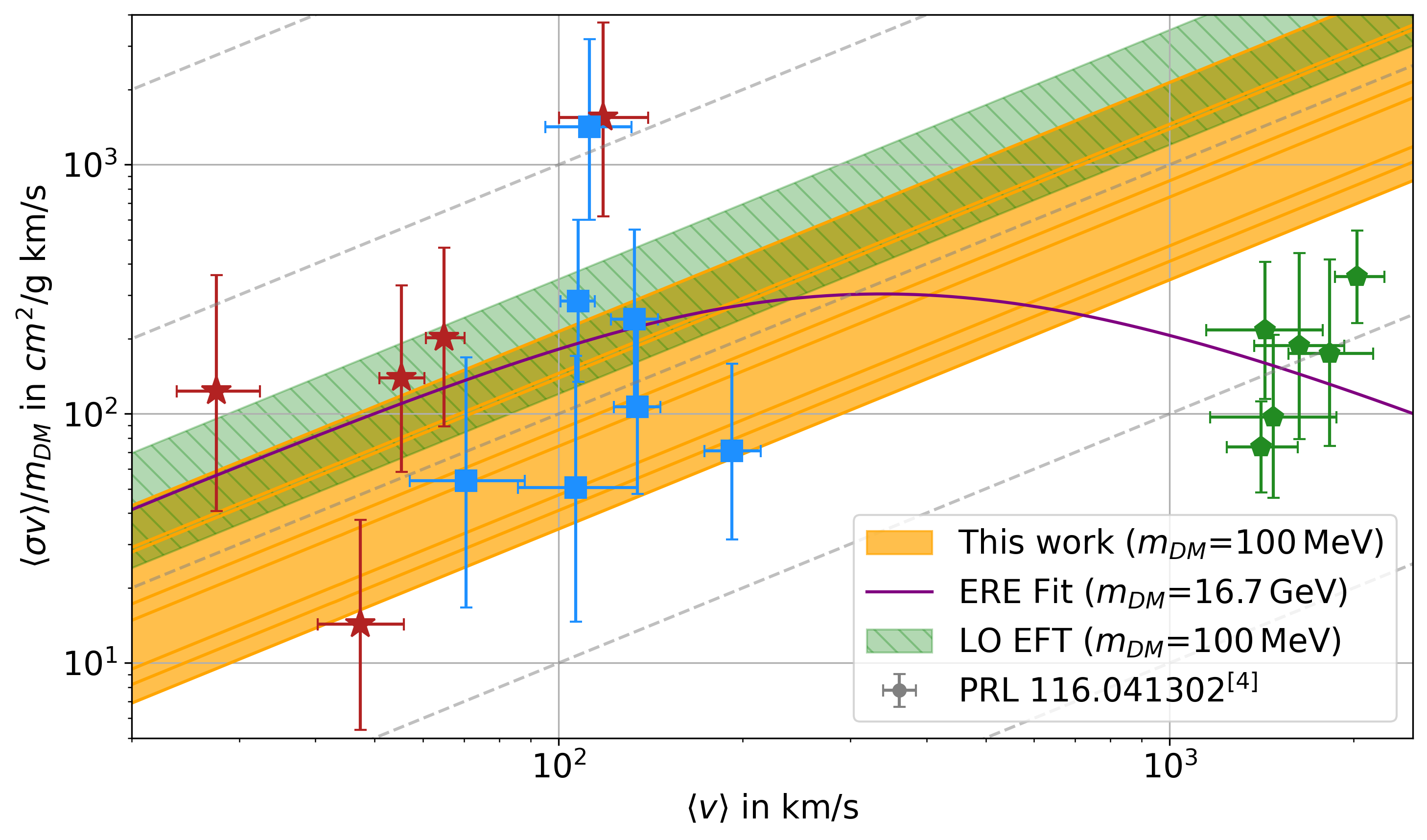}
    \caption{Comparison of the lattice data shown as orange lines with \cite{Kaplinghat:2015aga} represented by different symbols and colors that relate to different cosmological objects (red stars - dwarf galaxies, blue squares - low surface brightness, green pentagons - galaxy clusters). The y-axis is calculated with \eqref{eq: sigma v} and scales like $1/m_{\text{DM}}^3$. Lines of constant cross-section are shown in gray, dashed lines. The yellow band is an envelope over all of our ensembles shown in individual lines. The green hatched band shows the result using the scattering length predicted from $\chi$PT \cite{Bijnens:2011fm} given by \eqref{eq:Bijnens} using values of $4.6 \leq m^\infty_\pi/f_\pi \leq 6.0$ obtained from our ensembles. The effective range was set to zero in this case, which is motivated by the very low values for $r_0$ in the fit from \cite{Kondo:2022lgg} shown in purple.}
    \label{plot:sigma_v_v}
\end{center}
\end{figure}

\subsection{Energy levels}

We report the extracted energy levels and the renormalized pion decay constants in table~\ref{t:ensembles}. We further give the value of the average plaquette $\langle P \rangle$ since this quantity enters the estimation of the renormalization constant of $f_\pi$. 
The energies of the single-meson states are always smaller than unity in lattice units, i.e. $am_\pi < 1$, whereas the energy of the two-pion state below the first inelastic threshold is typically $aE_{\pi\pi} < \pi/2$ for all ensembles and $aE_{\pi\pi} < 1$ for most ensembles with $\beta > 6.9$. This suggests that discretization artifacts are not sizable, in particular for $\beta > 6.9$. We discuss this in appendix~\ref{s:systematics}, together with systematic effects due to the size of our (spatial) volumes. The results presented in the remainder of the main text can be considered free of any observed systematic effects within statistical errors. 

The energy shift between the interacting and non-interacting two-pion energies and its associated lattice momentum $p$ follows from the difference of twice the pion mass extrapolated to infinite volume and the two-pion energy on a lattice of finite volume according to Eq.~\eqref{eq:dispersion_relation}. This completes the set of quantities needed to determine the scattering properties.

\subsection{Scattering}
\begin{table}
	\centering
	\setlength{\tabcolsep}{2pt}
	\begin{tabular}{|c|c|c|c|c|}
		\hline
		$\beta$ & $a m_{0}$ & $a m_\pi^\infty\times 10^4$ & $a_0 m_\pi$ & $r_0 m_\pi$ \\ \hline \hline
		6.9 & -0.87 & $7401^{+8}_{-9}$ & $0.41^{+0.35}_{-0.27}$ & $50^{+363}_{-98}$\\
		6.9 & -0.9 & $5608^{+4}_{-4}$ & $0.42^{+0.07}_{-0.09}$ & $10^{+4}_{-2}$\\
		6.9 & -0.91 & $4845^{+9}_{-9}$ & $0.36^{+0.12}_{-0.11}$ & $43^{+51}_{-25}$\\
		6.9 & -0.92 & $3844^{+19}_{-30}$ & $0.52^{+0.24}_{-0.21}$ & $6.9^{+10.4}_{-4.1}$\\
		7.05 & -0.835 & $4373^{+9}_{-9}$ & $0.71^{+0.16}_{-0.26}$ & $1.9^{+1.4}_{-0.4}$\\
		7.05 & -0.85 & $3297^{+12}_{-13}$ & $0.56^{+0.14}_{-0.22}$ & $4.4^{+7.7}_{-2.1}$\\
		7.2 & -0.78 & $3696^{+4}_{-4}$ & $0.73^{+0.07}_{-0.08}$ & $2.2^{+0.3}_{-0.2}$\\
		7.2 & -0.794 & $2837^{+13}_{-14}$ & $0.88^{+0.13}_{-0.14}$ & $1.2^{+0.5}_{-0.4}$\\
	\hline
	\end{tabular}
	\caption{Results for the infinite volume pion mass calculated with \eqref{eq:infinite_volume} for all of our ensembles. We also report the results for the scattering length and the effective range from fitting \eqref{eq:ERT}. The large uncertainties for the effective range in some ensembles are unproblematic for the rest of the calculations as we are mainly interested in the low-energy behavior.}
	\label{t:results}
\end{table}

The Lüscher analysis grants access to the full scattering phase shift $\delta_0$. We report here results for the phase shift in the 14-dimensional irrep of two-pion scattering. We compare the resulting cross-section to astrophysical data.

The upper panel of figure~\ref{plot:p_cot_PS} shows the left-hand-side of \eqref{eq:ERT} for one of our ensembles ($\beta = 7.05,\, am_0 = -0.85$). As we do neither expect nor see any hints of resonances, and are interested in non-relativistic properties we use the ERE expansion \eqref{eq:ERT} to fit $a_0$ and $r_0$. The result of the fit is indicated with the blue line and band. For the calculation of the velocity-weighted cross-section \eqref{eq: sigma v}, we will use the results of the fit at non-relativistic velocities, where ERE is valid, because of $r_0 k \ll 1$. With \eqref{eq:ERT_sigma}, we can correlate these data points one-to-one to a cross-section and a center-of-momentum energy $s=E_{\pi\pi}^2$ (see lower panel of figure~\ref{plot:p_cot_PS}). At this point, we would like to remind the reader, that we are interested in the velocity-weighted cross-section \eqref{eq: sigma v} for which we need the dependence on the relative velocity which can be calculated from $s$ by
\begin{equation}
    v = 2\sqrt{1-\frac{4m^{\infty 2}_\pi}{s}}.
\end{equation}
\noindent The results of the fits for all of our ensembles is summarized in figure~\ref{plot:scattering_length_fpi} where we plot the scattering length against the value of $m_\pi^\infty/f_\pi$. Different data points correspond to different values of $\beta$ and $am_0$. We consistently observe a positive scattering length\footnote{The results are in qualitative agreement with those also observed in QCD (see e.g. \cite{Rodas:2023gma, RBC:2023xqv, Bruno:2023pde} for recent results) and $SU(2)$ \cite{Arthur:2014zda}, where the qualitative behavior is associated with appearance of Adler zeros. These are also expected in the present case.}. We also show the leading-order $\chi$PT prediction from \cite{Bijnens:2011fm} that is given by

\begin{equation}
    \label{eq:Bijnens}
    a_0 m^\infty_\pi = \frac{1}{32}\left(\frac{m^\infty_\pi}{f_\pi}\right)^2,
\end{equation}

\noindent in our sign convention. With that we can use our results to test the applicability of $\chi$PT. While the data points for $\beta$ = 7.05 and 7.2 agree with the $\chi$PT prediction within 1 to 2 $\sigma$, the data points for $\beta$ = 6.9 do not. This might be an effect of the limited amount of data points available or of the discretization of the lattice. On top of that one has to mention that the values of $m_\pi^\infty/f_\pi$ are large and one would naively not expect $\chi$PT to work in that regime. At present, we can neither confirm nor rule out the validity $\chi$PT in that regime from our results alone. See also \cite{Kulkarni:2022bvh,Bennett:2019jzz} on this question.

The plot also shows an estimate for the scattering length averaged over all ensembles as a gray band of $a_0m^\infty_\pi=0.57^{+0.25}_{-0.24}$. We can estimate the low energy cross-section from that by $\sigma_0\approx 4 \pi a_0^2$ and use it together with the constraint from astrophysical data\footnote{Due to the sizable uncertainties involved in the astrophysical data, we approximate them by $\sigma/m<0.2$ cm$^2$/g.} from \cite{Andrade:2020lqq,Eckert:2022qia,Sagunski:2020spe} to fix the lattice constant and thereby the mass of the dark matter candidate. We obtain a lower bound for the dark matter mass by of $m_{DM}\gtrsim110$ MeV, which fits the estimated mass range for SIMP dark matter based on perturbative considerations \cite{Hochberg:2014kqa} well.

Finally, in figure~\ref{plot:sigma_v_v}, we present our results for the velocity-weighted cross-section. The data points shown as circles are taken from \cite{Kaplinghat:2015aga}. Shown in a purple line are the fitted results from \cite{Kondo:2022lgg}. Their best estimate for the ERE are $a_0 = 22.2$ fm, $r_0 = -2.59\times 10^{-3}$ fm, and $m_{DM} = 16.72$ GeV. The green band shows the result using the scattering length predicted by leading-order $\chi$PT at values of $4.6<m^\infty_\pi/f_\pi<6.0$ which correspond to the minimal and maximal value obtained from our ensembles. $r_0$ was set to zero in that case, which we motivate by the small value of $r_0$ that was extracted from the data in \cite{Kondo:2022lgg}. Finally, the orange band shows an error estimate for our results from the lattice, where we used the ERE from every ensemble to calculate the velocity-weighted cross-section. All ensembles result in lines within that band. Our data show no velocity dependence in the regime relevant for dark matter in halos.

It should be noted that we can fix the lattice constant, and by that the dark matter mass, at any point during the calculation,
\begin{equation}
    \frac{\left<\sigma v\right>}{m_{DM}} \propto \left(\frac{1}{m_{DM}}\right)^3.
\end{equation}
\noindent This allows us to move the band up and down at will by changing the mass of the dark matter candidate. The mass of 100 MeV was chosen such that the band is compatible with the data.

\noindent The velocity, however, does not depend on the lattice constant. Therefore, any dynamics in the graph is fixed to the corresponding velocity. The velocity-weighted cross-section will decrease eventually due to the suppression by the velocity distribution for large, i.\ e.\ relativistic, velocities. We investigated this for the case of velocities much larger than $v_{esc}$ and find that the maximum lies at relativistic speeds. This discrepancy with the fitted curve can also be seen in the parameters of the ERE. Neglecting the effective range, the relevant quantity is $a_0 m_\pi$ which differs by more than three orders of magnitude ($\approx$ 1900 for \cite{Kondo:2022lgg} in natural units) between the results from the lattice and the best fit.

The data points are not direct measurements, but are based on the results of simulations that rely on assumptions about the dynamics of matter on galactic scales. Given the current uncertainties in the data and approximations employed here, this result does not alone invalidate dark Sp(4) theories within the fermion mass range investigated here.

\section{Discussion and conclusions}\label{s:conclusion}

Strongly-interacting models are a promising candidate for particle dark matter as they give solutions to known problems related to dark matter. In this paper, we used lattice field theory to study scattering properties of a specific realization of SIMP dark matter. Together with studies on the mass spectrum \cite{Kulkarni:2022bvh,Bennett:2017kga,Bennett:2023rsl,Bennett:2019jzz}, these non-perturbative results can also be used to determine low energy constants in an effective description with $\chi$PT \cite{Kulkarni:2022bvh,Bennett:2017kga}. 

These results are a first step towards a robust determination of the scattering properties of composite dark matter models. The analysis can straightforwardly be improved by using finer lattices, while ensuring that $aE_{\pi\pi}$ remains smaller than unity in all cases. This will further reduce the influence of existing finite lattice spacing effects. More energy levels can be obtained on the lattice by enlarging the operator basis, e.\ g.\ with smeared interpolating operators and correlators with non-vanishing momentum. On the smaller lattices subleading finite volume corrections proportional to $e^{-m_\pi L}$ \cite{Bedaque:2006yi} can be taken into account.

We find a rough lower bound for the mass of the dark pions, assuming that the isospin-2 channel is dominant, of order 100 MeV. This also allows to at least partially describe first exploratory results on the self-interaction rates of dark matter from astrophysical data. Further research is needed to provide a stronger statement on the validity of this model as a dark matter candidate. This includes both the technical aspects mentioned in the previous paragraph and the inclusion of other scattering channels.

Going beyond the isospin-2 scattering channel, the natural next step is the determination of the phase shift in the isospin-1 channel including the $3\to2$ process as well as further derived quantities. 

\acknowledgments
We thank Elizabeth Dobson, Suchita Kulkarni and Jong-Wan Lee for a critical reading of the manuscript and helpful comments. We especially thank Suchita Kulkarni for the helpful discussions. We thank the authors of \cite{Bennett:2019jzz} for sharing their gauge configurations for the purposes of this paper. YD and FZ have been supported the Austrian Science Fund research teams grant STRONG-DM (FG1). FZ has been supported by the STFC Grant No. ST/X000648/1. The computations have been performed on the Vienna Scientific Cluster (VSC4).  

{\bf Research Data Access Statement}---The data generated and the analysis code for this manuscript can be downloaded from  Ref.~\cite{data_release, analysis_release}. 

{\bf Open Access Statement}---For the purpose of open access, the authors have applied a Creative Commons 
Attribution (CC BY) license to any Author Accepted Manuscript version arising.

\appendix

\section{Lattice systematics}\label{s:systematics}

\begin{figure}
\begin{center}
    \includegraphics[width=0.49\textwidth]{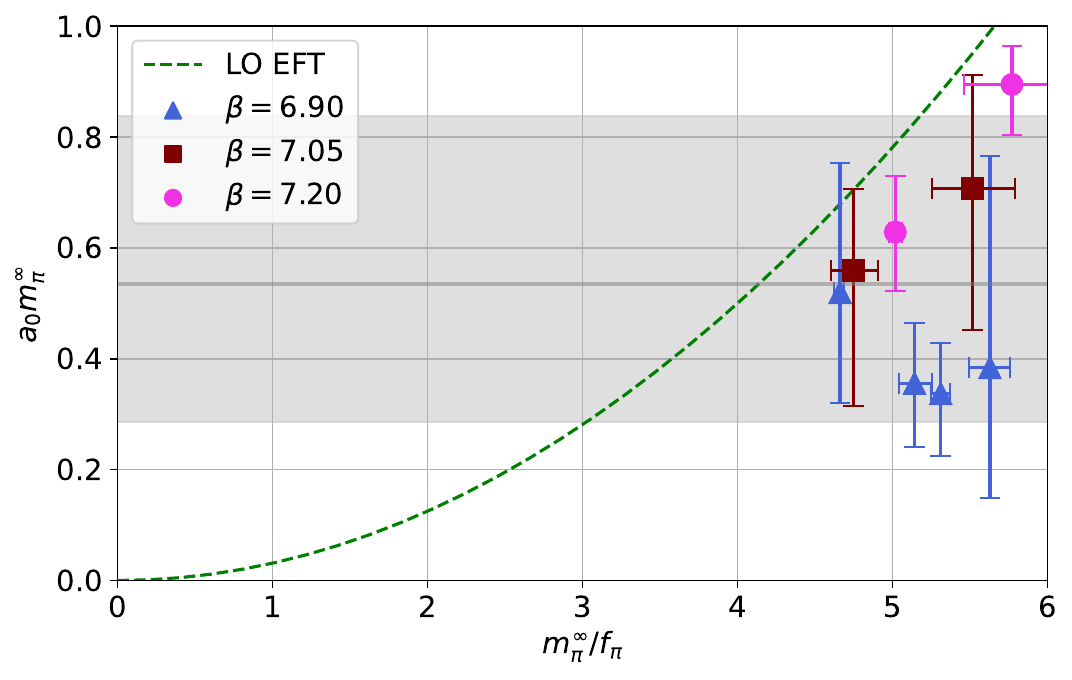}
    \includegraphics[width=0.49\textwidth]{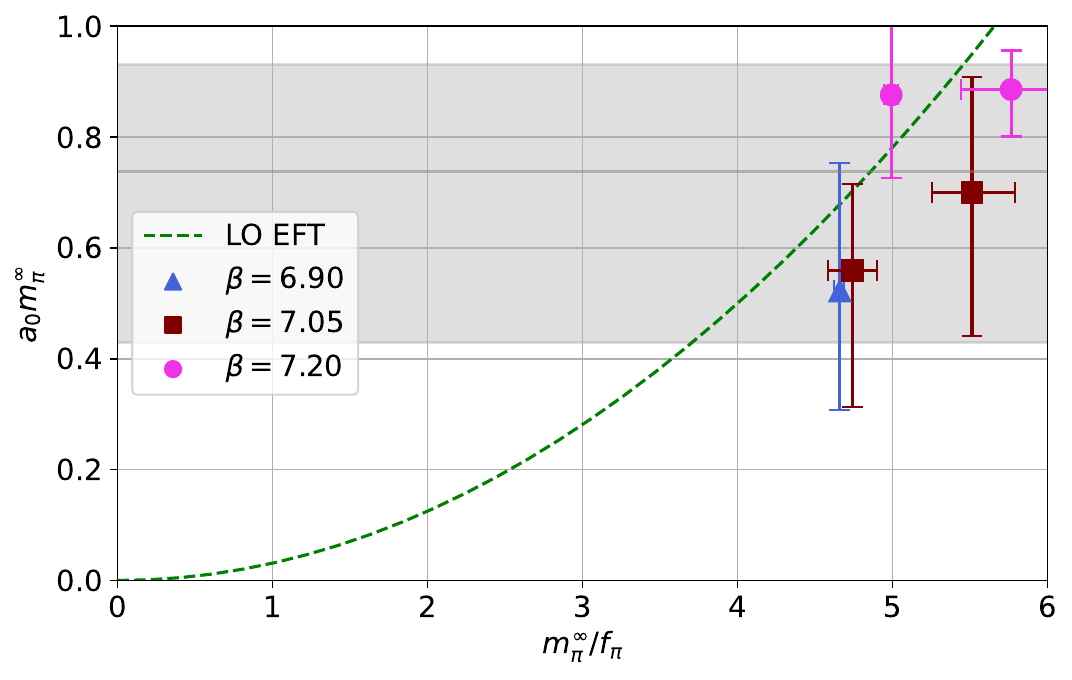}
    \includegraphics[width=0.49\textwidth]{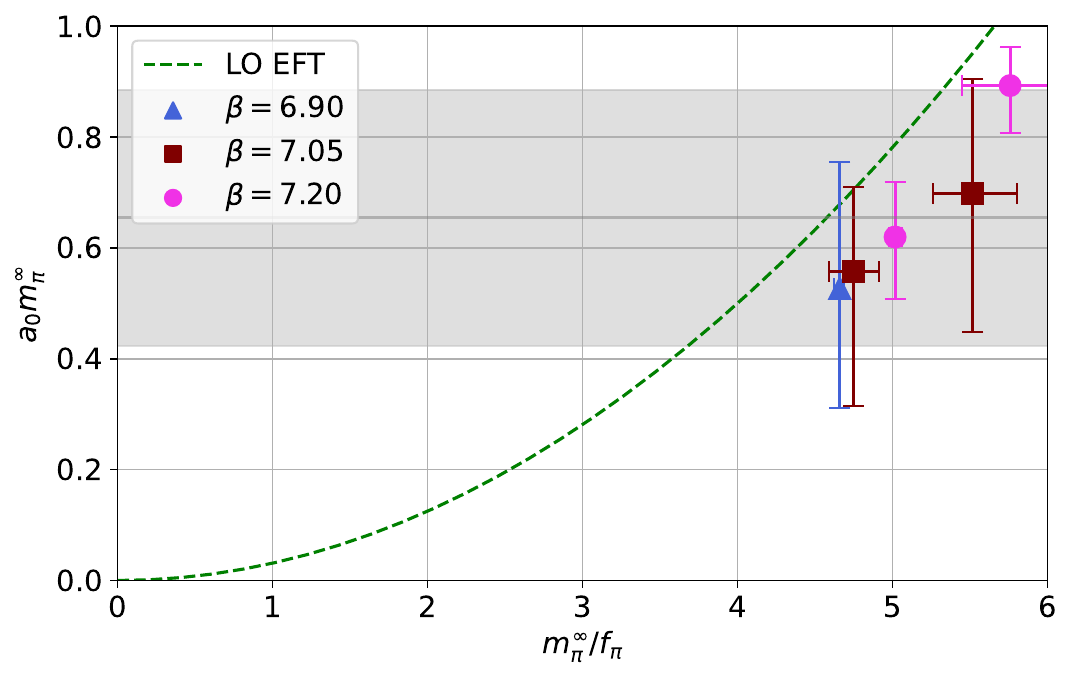}
    \caption{Same as figure \ref{plot:scattering_length_fpi}, but restricted to selected ensembles to study the lattice artifacts. (top left) All ensembles with $N_L=8$ and $m_\pi / m_\pi^{\infty} > 1.3$ have been excluded. (top right) All ensembles with $a E_{\pi\pi} < 0.95$ have been excluded. (bottom) The two restrictions have been combined.}
    \label{plot:lattice_artifacts}
\end{center}
\end{figure}

In order to assess the relevance of lattice artifacts we repeated the analysis, by excluding select ensembles from Tab.~\ref{t:ensembles}. We show the results in figure \ref{plot:lattice_artifacts}. First, we restricted ourselves to ensembles where $a E_{\pi\pi} \lesssim 1$ where discretization artifacts are expected to be smaller. In particular, we choose $a E_{\pi\pi} < 0.95$ to exclude the ensemble with $\beta=6.9$ and $am_0=-0.91$ where all two-pion energies are very close to one. Secondly, we also redid the analysis by excluding the smallest lattices with $N_L=8$ and all ensembles where $m_\pi / m_\pi^{\infty} > 1.3$. Finally, we combined the two exclusion criteria. 

The omission of smaller lattices and ensembles with sizable finite volume effects on the pion mass leads to only minor changes, which are compatible with statistical uncertainties. After restricting the analysis to $a E_{\pi\pi} < 0.95$, we find that our results are in better agreement with leading-order chiral perturbation theory. When combining the two restriction our data reflects the expected mass-dependence within the statistical uncertainties. 

\begin{figure}
    \centering
    \includegraphics[width=0.49\linewidth]{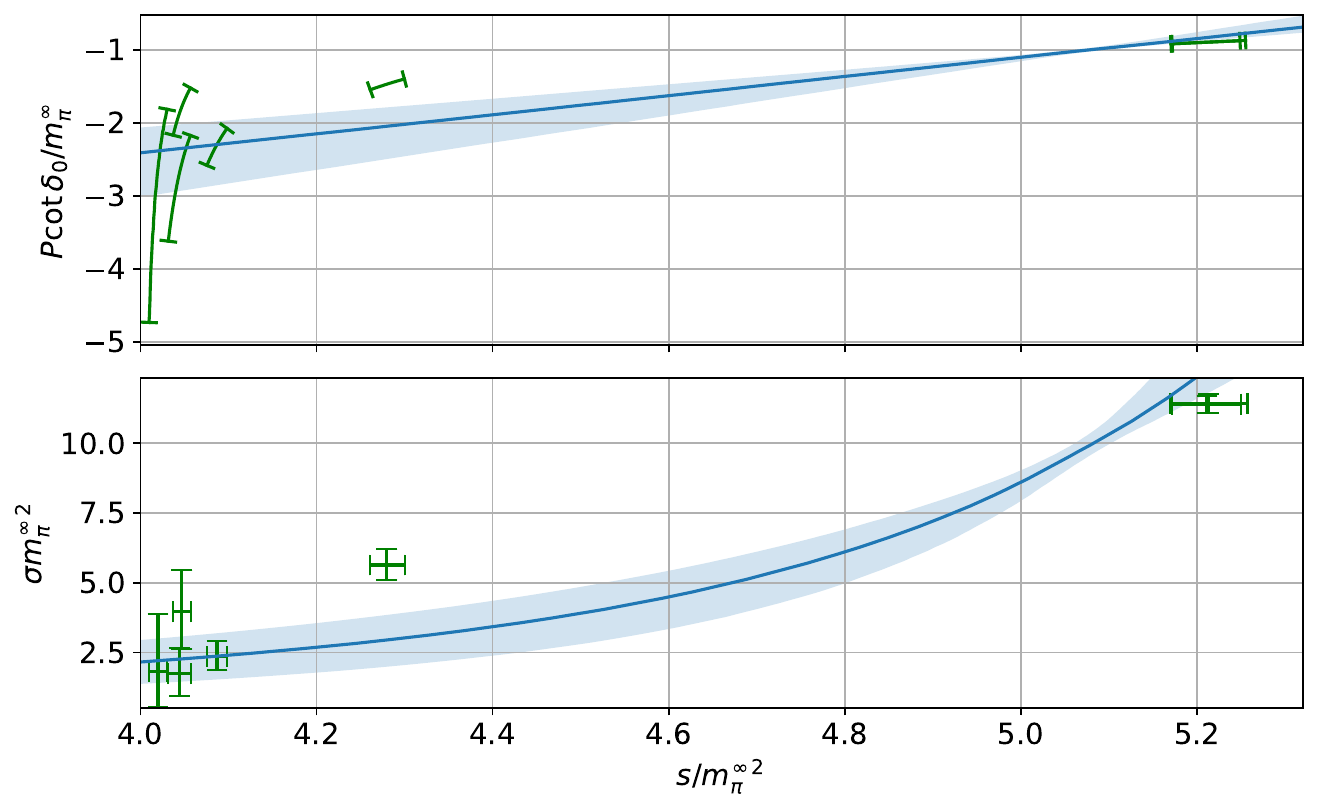}
    \includegraphics[width=0.49\linewidth]{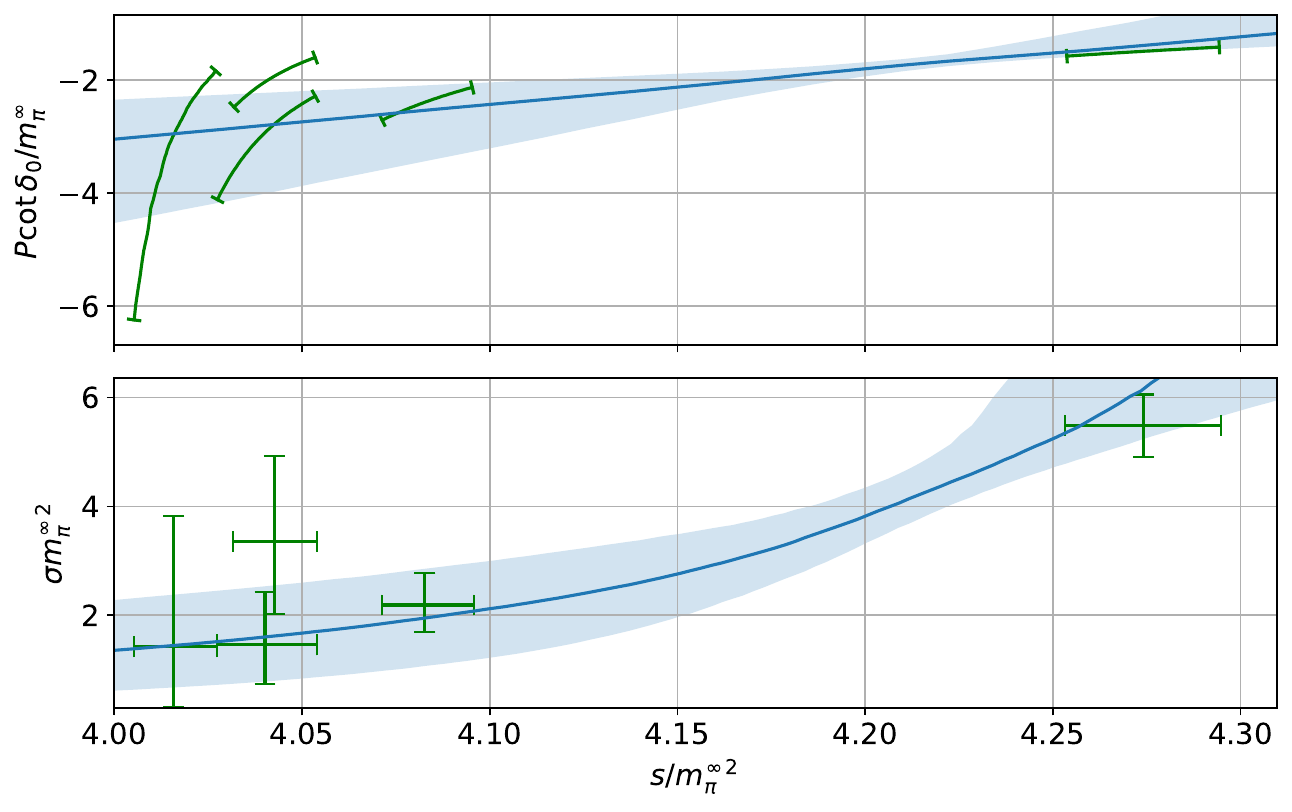}
    \caption{Fit of the discrete data on the phase shift to the ERE for the ensembles with $\beta=6.9$ and $m_0=-0.90$. (left) Fit using all ensembles listed in table~\ref{t:ensembles}. The ensembles with $L=10$ is in noticeable tension to the ERE fit. (right) The same fit where the ensembles with $L=8$ have been excluded. No tension for the $L=10$ is observed.}
    \label{fig:ERE_systematics}
\end{figure}

Additionally, we discuss the fit of the phase shifts calculated according to Eq.~\ref{eq:tan_PS} to the ERE in Eq.~\ref{eq:ERT} and the resulting scattering cross-section in Eq.~\ref{eq:ERT_sigma}. In some cases, the resulting fits were in tension with some data points of the calculated phase shift by up to three standard deviations. This might highlight that the phase shift at the largest values of $s/{m_\pi^\infty}^2$ is not well described by the ERE. In our approach a large value of $s/{m_\pi^\infty}^2$ corresponds to a smaller lattice and potentially larger discretization artifacts. The apparent tension in the fit disappears if the data point at the largest value of $s/{m_\pi^\infty}^2$ is excluded from the fit. We show an example of this in Fig.~\ref{fig:ERE_systematics}. 

In table~\ref{t:ERE_systematics}, we compare the values of $a_0$ and $r_0$ for various restrictions in the analysis.  

\begin{table}
    \centering
    \begin{tabular}{|c|c|c|c|c|c|c||c|c|c|c|c|}
         \hline 
         $\beta$ & $am_0$ & $a_0 m_\pi^\infty (a)$ & $a_0 m_\pi^\infty (b)$ & $a_0 m_\pi^\infty (c)$ & $a_0 m_\pi^\infty (d)$ & $a_0 m_\pi^\infty (e)$ & $r_0 m_\pi^\infty (a)$ & $r_0 m_\pi^\infty (b)$ & $r_0 m_\pi^\infty (c)$ & $r_0 m_\pi^\infty (d)$ & $r_0 m_\pi^\infty (e)$ \\
         \hline \hline 
            6.9     & -0.87  & $0.39^{+0.35}_{-0.25}$ & same as $(a)$ & - & - & same as $(a)$ & $59^{+329}_{-110}$ & same as $(a)$ & - & - & same as $(a)$ \\ 
            6.9     & -0.90  & $0.45^{+0.06}_{-0.07}$ & $0.33^{+0.09}_{-0.11}$ & - & - & same as $(b)$ & $9.3^{+3.5}_{-2.2}$ & $47^{+59}_{-23}$ & - & - & same as $(b)$ \\ 
            6.9     & -0.91  & $0.36^{+0.12}_{-0.12}$ & same as $(a)$ & - & - & same as $(a)$ & $42^{+57}_{-27}$ & & - & - & same as $(a)$ \\ 
            6.9     & -0.92  & $0.52^{+0.23}_{-0.21}$ & same as $(a)$ & same as $(a)$ & same as $(a)$ & same as $(a)$ & $6.7^{+9.5}_{-3.8}$ & same as $(a)$ & same as $(a)$ & same as $(a)$ & same as $(a)$ \\ 
            7.05    & -0.835 & $0.70^{+0.12}_{-0.21}$ & $0.70^{+0.21}_{-0.26}$& same as $(b)$ & same as $(b)$ & same as $(b)$ & $1.9^{+1.1}_{-0.4}$ & $6.4^{+12.0}_{-4.0}$ & same as $(b)$ & same as $(b)$ & same as $(b)$  \\ 
            7.05    & -0.85  & $0.60^{+0.14}_{-0.25}$ & same as $(a)$ & same as $(a)$ & same as $(a)$ & same as $(a)$ & $3.7^{+7.4}_{-1.7}$ & same as $(a)$ & same as $(a)$ & same as $(a)$ & same as $(a)$ \\ 
            7.2     & -0.78  & $0.80^{+0.06}_{-0.08}$ & $0.88^{+0.08}_{-0.08}$ & same as $(b)$ & same as $(b)$ & $0.88^{+0.07}_{-0.08}$ & $2.0^{+0.3}_{-0.2}$ & $2.1^{+0.6}_{-0.4}$ & same as $(b)$ & same as $(b)$ & $2.1^{+0.6}_{-0.4}$  \\ 
            7.2     & -0.794 & $0.84^{+0.13}_{-0.16}$ & $0.62^{+0.10}_{-0.11}$ & same as $(a)$ & same as $(b)$ & same as $(a)$ & $1.3^{+0.7}_{-0.4}$ & $3.7^{+1.9}_{-1.1}$ & same as $(a)$ & same as $(b)$ & same as $(a)$ \\ 
         \hline
    \end{tabular}
    \caption{Results of the ERE fit parameters $a_0$ and $r_0$ in units of the pion mass using different exclusion criteria. (a) All ensembles in table~\ref{t:ensembles} have been included. (b) All ensembles with $N_L>8$ and $m_\pi / m_\pi^{\infty} < 1.3$ (c) All ensembles with $E_{\pi\pi} < 0.95$ (d) All ensembles with $N_L>8$, $m_\pi / m_\pi^{\infty} < 1.3$ and $E_{\pi\pi} < 0.95$ (e) The smallest lattices have been excluded such that the ERE fits the data compatible within one standard deviation.}
    \label{t:ERE_systematics}
\end{table}

Overall, we find no major deviations in our final results for the scattering length $a_0$ due to lattice artifacts. In particular, our range of $a_0 m_\pi^\infty$ remains virtually unchanged -- see table \ref{t:ERE_systematics}.

\bibliography{bibliography.bib}

\end{document}